\documentclass[twoside]{article}
\usepackage[dvips]{graphicx}
\usepackage{fleqn,espcrc2}

\usepackage[figuresright]{rotating}

\newcommand{\be}{\begin{equation}}
\newcommand{\ee}{\end{equation}}
\newcommand{\bc}{\begin{center}}
\newcommand{\ec}{\end{center}}
\newcommand{\bes}{\begin{equation*}}
\newcommand{\ees}{\end{equation*}}
\newcommand{\beqn}{\begin{eqnarray}}
\newcommand{\eeqn}{\end{eqnarray}}
\newcommand{\beqns}{\begin{eqnarray*}}
\newcommand{\eeqns}{\end{eqnarray*}}
\newcommand{\bs}{B_s^0\!-\!\bar B_s^0}
\newcommand{\widdif}{\frac{\Delta \Gamma_{B_s}}{\Gamma_{B_s}}}
\newcommand{\barms}{\overline{\mathrm{MS}}}
\newcommand{\err}[2]{\raisebox{0.08em}{\scriptsize {$\;\begin{array}{@{}l@{}}
			  \plus\makebox[2.05em][r]{#1} \\[-0.12em] 
			  \minus\makebox[2.05em][r]{#2} 
			\end{array}$}}}

\newcommand{\errd}[2]{\raisebox{0.08em}{\scriptsize {$\;\begin{array}{@{}l@{}}
			  \plus\makebox[1.35em][r]{#1} \\[-0.112em] 
			  \minus\makebox[1.35em][r]{#2} 
			\end{array}$}}}
\newcommand{\plus}{\makebox[2pt][c]{$+$}}
\newcommand{\minus}{\makebox[2pt][c]{$-$}}

\newcommand{\AmS}{{\protect\the\textfont2
  A\kern-.1667em\lower.5ex\hbox{M}\kern-.125emS}}

\title{Quenched and first unquenched lattice HQET determination of the
$B_s$-Meson width difference.}

\author{V. Gim\'enez\address{Dep. de F\'\i sica Te\'orica, IFIC and Univ. de
Valencia,\\  
Dr. Moliner 50, E-46100, Burjassot, Valencia, Spain} and J. Reyes\addressmark
\thanks{Govierno Vasco fellowship.}\thanks{Talk presented by J. Reyes.}}
\begin{document}

\begin{abstract}
We present recent results for the prediction of the $\bs$ lifetime difference
from lattice Heavy Quark Effective Theory simulations. In order to get a
next-to-leading order result  we have calculated the matching between QCD
and HQET and the two-loop anomalous dimensions in the HQET for all the $\Delta B\!\!=\!\!2$
operators, in particular for the operators which enter the width difference.
We present results from quenched and, for the first time, from unquenched
simulations.
We obtain for the $\bs$ lifetime difference, $\widdif^{(que.)}=(5.1\pm 1.9\pm 1.7)\times
10^{-2}$ and  $\widdif^{(unq.)}=(4.3\pm 2.0\pm 1.9)\times
10^{-2}$ from the quenched and unquenched  simulations respectively.
\vspace{1pc}
\end{abstract}

\maketitle

\section{Introduction.}
In the Standard Model, the width difference ($\widdif$) in the $\bs$ system is
expected to be the largest among bottom hadrons and could be measured in the
near future. The Standard Model prediction for $\widdif$
relies on an operator product expansion, where the short distance 
scale is the $b$ quark mass \cite{ben97}. The theoretical expression
for the width difference reads, schematically \cite{APE00},
\be\label{dgg}
\widdif=
K\left(G(z)
    +G_S(z){\cal R}(m_b)+
   {\delta}_{1/m}\right)\,
\ee
where $K$ is a factor which encloses known parameters, 
$z=m_c^2/m_b^2$, $G(z)$ and $G_S(z)$ are NLO Wilson
coefficients calculated in 
\cite{ben98} and $\delta_{1/m_b}$ are $O(1/m_b)$ corrections which depend on matrix
elements of higher dimension operators (see \cite{ben97} for details). All the non-perturbative
contribution, at leading order in $1/m_b$, comes 
from $\cal R$,
\be
{\cal R}(m_b)\equiv \frac{\langle\bar B_s|{O_S(m_b)}|B_s\rangle}
{\langle\bar B_s|{O_L(m_b)}|B_s\rangle},
\ee
where 
\beqn
O_L&=&\bar b\, \gamma^{\mu}\, (1 -\gamma_{5})\, s\,  
\, \bar b\, \gamma_{\mu}\, (1 -\gamma_{5})\, s\nonumber\\
O_S &=& \bar b\, (1-\gamma_{5})\, s\, 
 \bar b\,  (1-\gamma_{5})\, s\label{OL-OS}
 \eeqn
We present results for the ${\cal R}$ parameter and for other $\Delta B\!\!=\!\!2$
B-parameters from lattice HQET in the
quenched approximation and also, for the first time, from unquenched
simulations. 
We have performed the matching between lattice HQET and continuum QCD at NLO.
Until now this matching  was known at NLO only for the $O_L$ operator (see
ref.~\cite{vicent97}). We have extended this calculation to all $\Delta B\!\!=\!\!2$
operators by computing 
the NLO matching
between QCD and HQET in the continuum and the relevant two-loop
 anomalous dimensions \cite{our2}.
\section{B-parameters from the lattice.}
In the HQET there are four independent operators which we write in terms of the
B-parameters, defined as vacuum insertion deviations:
\begin{eqnarray}\label{qb}
\frac{\langle\bar B_{d(s})|O_L(\mu)|B_{d(s})\rangle}{\frac{8}{3}f^2_{B_{d(s})}M^2_{B_{d(s})}} &\equiv&  B(\mu), \\
\label{qsbs}
\frac{\langle\bar B_{d(s})|O_S(\mu)|B_{d(s})\rangle}{-\frac{5}{3}f^2_{B_{d(s})}M^2_{B_{d(s})}{\cal
X}} \!\!\!&\equiv&\!\!\!  B_S(\mu)\\ 
\frac{\langle\bar B_{d(s})|O^{LR}(\mu)|B_{d(s})\rangle}
{ -2f^2_{B_{d(s})}M^2_{B_{d(s})}}
\left(1+\frac{2}{3}{\cal X}\right)^{\!\!\!\!-1} \!\!\!&\equiv&\!\!\!\! B^{LR}(\mu) \\
\frac{\langle\bar B_{d(s})|O_S^{LR}(\mu)|B_{d(s})\rangle}
{\frac{1}{3}f^2_{B_{d(s})}M^2_{B_{d(s})}}
\left(1+6\,{\cal X}\right)^{-1} \!\!\!&\equiv&\!\!\!\!  B_S^{LR}(\mu)
\end{eqnarray}
where,
\be\label{Xi}
{\cal X}\equiv\frac{M_{B_{d(s)}}^2}{(m_b+m_{d(s)})^2}\qquad ,
\ee
$O_L$ and $O_S$ are defined in eq. (\ref{OL-OS}) and
\beqn
O^{LR}&=&\bar b\, \gamma^{\mu}\, (1 -\gamma_{5})\, q\,  
\, \bar b\, \gamma_{\mu}\, (1 +\gamma_{5})\, q,\\
O_S^{LR}&=& \bar b\, (1-\gamma_{5})\, q\, 
 \bar b\,  (1+\gamma_{5})\, q.
 \eeqn

The lattice B-parameters are extracted from the large time behavior
of the ratio between three- and two-point correlation functions (see
\cite{gim&mar96}):
\[
R_{O_i}\equiv\frac{C_{O_i}(-t_1,t_2)}{C(-t_1)C(t_2)}\stackrel{t_1,t_2\to
\infty}{\longrightarrow} 
\frac{\langle\bar B_{s}|O_i(a)|B_{s}\rangle}{|\langle
0| A_0(a)|B_{s}\rangle|^2}
\]
where
\beqns
C_{O_i}(t_1,t_2)\!\!\!\!\!&\equiv&\!\!\!\!\!\!\!\!\sum_{\vec{x}_1,\vec{x}_2} \langle
0|A_0(\vec{x}_1,t_1) O_i(\vec{0},0) A^{\dagger}_0(\vec{x}_2,t_2)|0\rangle\\
C(t)&\equiv&\sum_{\vec{x}} \langle
0| A_0(\vec{x},t) A^{\dagger}_0(\vec{0},0)|0\rangle
\eeqns
and $A_{\mu}$ is the HQET axial current.

The results presented here are obtained from two simulations: a quenched one performed by
APE coll.~with the SW-Clover action on a $24^3\times 40$ lattice with 600 gauge
configurations at
$\beta=6.0$ with $a^{-1}=2$ $\mathrm{GeV}$ and an unquenched one carried out by
T$\chi$L coll.~(gauge configurations) and APE (correlation functions) with the
Wilson action on the 
same volume with 60 gauge configurations at
$\beta=5.6$ with two degenerate sea quarks corresponding to $a^{-1}=2.51,\,
2.54\, \mathrm{GeV}$. See refs.~\cite{gim&mar96} and \cite{TxL98}
respectively for details. 

\begin{figure}
\includegraphics[bb=1.5cm 1.5cm 17.5cm
18cm,scale=.3,angle=-90]{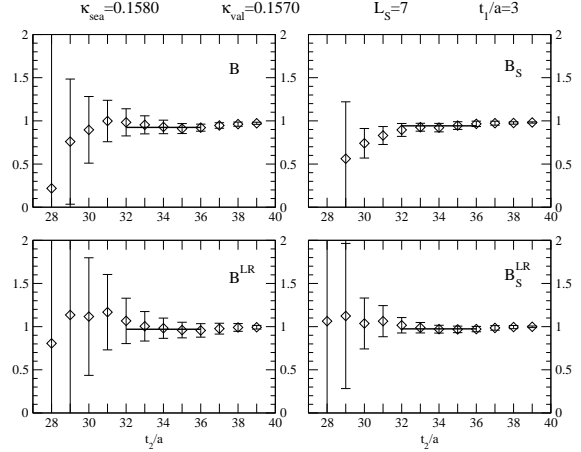}        
\caption{The ratios $R_{O_i}$ for different operators computed at $k_{val}=0.1570$
and $k_{sea}=0.1580$ for $t_1/a=3$ and with a smearing size of 7.}
\end{figure}
In figure 1 we show the plateaux for the four ratios $R_{O_i}$ in the unquenched
case. We observe good plateaux over
large time-distances for all operators (In ref. \cite{vicent97}, the
corresponding figures in the quenched case were presented).

In \cite{beach} we pointed out that different groups using different methods found
in quenched simulations
that  the lattice B-parameters in the static
limit are very close to 1. As one can see from figure
1, this result holds also in unquenched simulations.

\section{$\mathbf{1/m_b}$ dependence of the B-parameters.}
Having performed the calculation described in the previous section one obtains
the B-parameters in QCD to leading order in $1/m_b$. 
By using
vacuum insertion approximation (VIA) for the subleading operators
\cite{mannel} and the fact that the bare
lattice 
B-parameters are very close to 1, we have shown \cite{our1} that all
$\Delta B\!\!=\!\!2$ B-parameters, defined as vacuum 
insertion deviations, have small $O(\bar
\Lambda/m_b)$ corrections ($\bar \Lambda \equiv M_{B}-m_b$), 
\be\label{mbexp}
B_i(m_b)=\bar B_i(m_b) + \underbrace{O(0.3\frac{\bar \Lambda}{m_b})}_{\sim
0.05}+O(\frac{1}{m_b^2}),
\ee
where $B_i(m_b)$ and $\bar B_i(m_b)$ are the full and the static QCD
B-parameters, respectively. The factor $0.3$ is an estimation of the deviation
from VIA in the calculation of the subleading operators and
$O(\alpha_s$) corrections.  
\section{Results.}
Before presenting our results, let us stress that $m_s$ and $m_b$ parameters in 
previous sections are the corresponding pole masses. The latter coincides with 
the expansion parameter of the HQET because we have set the  residual mass to
zero. 
We calculate the $b$ quark pole mass from the running $\overline{\mathrm{MS}}$
mass, which can be accurately determined. Since our computation is
performed at NLO, we use the perturbative relation between the pole and the
running mass at the same order. From the world average running
mass \cite{gim00}, $\overline m_b(\overline m_b)=4.23\pm 0.07\,\,\mbox{GeV}$ one
obtains $m_b=4.6\pm 0.1\,\mbox{GeV}$. Notice that the contribution of $m_s$ is
very small because it always appears divided by $m_b$. 

\begin{table}[t]
\parbox{7cm}{\caption{Values of the B-parameters in the
NDR-$\overline{\mathrm{MS}}$ scheme (see text).}\label{tabl1}}\\
\begin{tabular}{ccc}
\hline 
B-Parameters&\parbox{2cm}{Clover quenched}&\parbox{2cm}{Wilson unquenched}\\
\hline 
\hline 
$B(m_b)$&0.83(5)(4)(5)&0.74(3)(8)(5)\\
$B_S(m_b)$&0.96(8)(5)(5)&0.87(2)(11)(5)\\
$B^{LR}(m_b)$&0.94(5)(4)(5)&0.98(4)(6)(5)\\
$B_S^{LR}(m_b)$&1.03(3)(5)(5)&1.10(2)(9)(5)\\
\hline 
\hline 
\end{tabular}
\end{table}

In table \ref{tabl1} we present the
values for all $\Delta B\!\!=\!\!2$ B-parameters. The first error in table 1 comes from lattice
simulations and includes statistical and systematic errors. The second one is an
estimate of the error due to the uncertainties in the values of the lattice
coupling constant and to higher order contributions to the matching. The third
is an estimate of $1/m_b$ corrections to the static result (see eq.(\ref{mbexp})).
The 
analysis of the 
perturbative matching is explained in detail in our previous work
\cite{nos99} where we gave
 $B(m_b)^{(quenched)}=0.81\pm 0.05\pm 0.04$. The tiny difference is due to the fact that
 there we used, in the perturbative  
evolution, a number of flavours $n_f=4$ instead of $n_f=0$ as in the
present paper. From table 1 we see that the systematic errors are larger in the
unquenched case than in the quenched case. This is due to the fact that the
renormalization constants for the Wilson action are bigger than those for the
SW-Clover one, and consequently, the contribution of higher orders in $\alpha_s$
are more important. Taking into account the errors in table 1 we cannot
establish any clear unquenched deviation from the quenched result.

The B-parameters in table \ref{tabl1} are QCD scheme dependent,
they all are in the NDR-$\barms$ scheme. To make 
contact with the computation of \cite{ben98}, we have subtracted the
evanescent operators in the renormalization of $O_L$ and $O_S$ as in \cite{ben98}.
For $B^{LR}$ and  $B_S^{LR}$ we use the prescription of \cite{buras00}.

For $\cal R$ (see eq.(2)), which contains all the non-perturbative
contribution to the width difference, 
at leading order in $1/m_b$,
we obtain
\beqn
{\cal R}(m_b)^{(que.)}&=&-0.95(7)(9)\\
{\cal R}(m_b)^{(unq.)}&=&-0.97(5)(15)
\eeqn
where as usual, the first error comes from lattice simulations, and the second is
systematic due to the uncertainty in the perturbative matching and to
the contribution of higher orders in $1/m_b$.
\begin{table} 
\parbox{7cm}{\caption{Values of  $B$, $B_S$ and ${\cal R}$ from different
groups.}\label{tabl2}}\\ 

\scalebox{.819}[.9]{\begin{tabular}{cccc}
\hline
\hline
\mbox{\rm Group}&\mbox{$B(m_b)$}&\mbox{$B_S(m_b)$}&\mbox{${\cal R}(m_b)$}
\\
\hline
\cite{JLQCD} &
$ 0.85(3)(11)$&$0.80(2)(10)$&$-0.91(5)(17)$
\nonumber\\\hline
\cite{APE00}
&$ 0.91(3)\err{0.00}{0.06}$&
$ 0.86(2)\err{0.02}{0.03}$&
$ -0.93(3)\err{0.00}{0.06}$\nonumber\\\hline
\parbox{1.6cm}{\mbox{This work} quenched} &$ 0.83(5)(6)$&$ 0.81(7)(6)$&
$ -0.95(7)(9)$\\
\hline
\parbox{1.82cm}{\mbox{This work} unquenched}&$ 0.74(3)(9)$&$ 0.74(2)(10)$&
$ -0.97(5)(15)$\\\hline
\hline
\end{tabular}}
\end{table}
\section{Comparison with other recent results}
In table \ref{tabl2} we present our result for the B-parameters relevant for
$\widdif$ compared with other recent quenched determinations. In \cite{APE00} the $B_S$
is defined as in eq.(\ref{qsbs}) but in terms of
the running mass instead of the pole mass. Therefore, we have multiplied our
value by $(\overline m_b(\overline m_b)/m_b)^2$ in order to compare with their
result. 
On the other hand, in the definition used in \cite{JLQCD} does not appear the
factor 
${\cal X}$ (see eq.~(\ref{Xi})), therefore we have divided their result
 by
${\cal X}\,(m_b/\overline m_b(\overline m_b))^2$. Notice that in spite of using different methods to obtain the
B-parameters, there is a good agreement between the three quenched
computations, in
particular in the value of the ratio ${\cal R}$. In the unquenched case the
central values of $B$ and $B_S$ are lower but still compatible within errors.
Nevertheless, the ratio ${\cal R}$, is in very good agreement with the quenched
computations.  

\begin{figure}[!t]
\includegraphics[bb=1.5cm 5cm 17.5cm 18.5cm,scale=0.28]{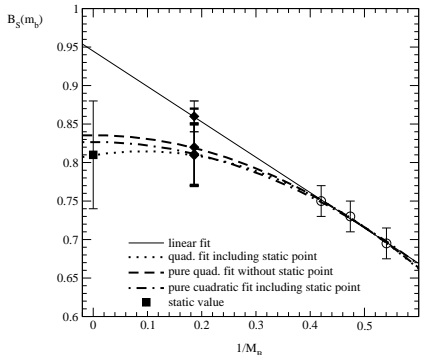}	
\caption{Linear and quadratics fits of $B_S(m_b)$ in $1/M_B$ to the ref.
 \protect{\cite{APE00}} data (open circles) combined with the static point (filled
square). The filled diamonds correspond to physical $B_s$ meson mass.}
\end{figure}
From equation (\ref{dgg}) we get our prediction,
\beqn
\widdif^{(que.)}&=&(5.1\pm 1.9\pm 1.7)\times 10^{-2}\\
\widdif^{(unq.)}&=&(4.3\pm 2.0\pm 1.9)\times 10^{-2}.
\eeqn
The first error is systematic obtained from the spread of values of
all input parameters in eq.(\ref{dgg}). The second one comes from the
uncertainty in the value of $ \delta_{1/m_b}$. Since in the estimate of
$ \delta_{1/m_b}$ the operator matrix elements were computed using VIA
 and the radiative corrections were not included, we assume an error of $30\%$
 \cite{APE00}. As can be seen, this parameter is still affected by a large
 uncertainty, so that a precise determination of the width difference requires
 the computation of the subleading matrix elements using lattice QCD. This
 simulation is missing to date.

Our result is to be compared with the present experimental status \cite{schn}
\be
\left(\frac{\Delta\Gamma_{B_s}}{\Gamma_{B_s}}\right)^{\mbox{exp.}}=\left(17\errd{09}{10}\right)\times
10^{-2}
\ee
As can be seen the central values are rather different but still compatible
within the large errors. 

\begin{figure}[t!]
\includegraphics[bb=1.5cm 1.5cm 14.6cm 18cm,scale=.3,angle=-90]{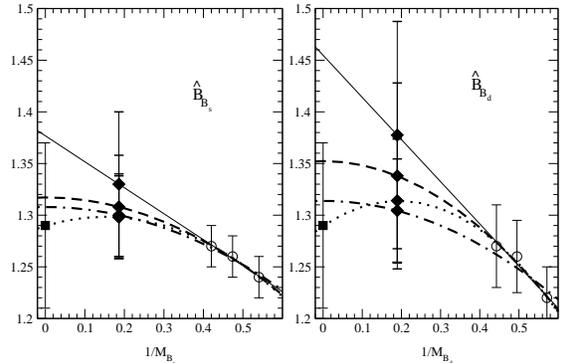}        
\caption{The same of figure 2 for $\hat B_{B_s}$ and $\hat B_{B_s}$ with the
data of ref. \protect{\cite{APE00b}}.}
\end{figure}
Finally, in figures 2 and 3 we present different fits in $1/M_B$ to the quenched
data of
refs.~\cite{APE00,APE00b} for $B_S$ and the renormalization
group invariant parameters $\hat B_{B_d}$ and $\hat B_{B_s}$. As we 
pointed out
in sec.~3, the HQET predicts that the $1/M_B$ correction to the static
B-parameters is small and the contribution of the quadratic term 
cannot be neglected. To study this dependence, we perform different
fits: a linear fit, a quadratic fit, including the linear term, and a pure
quadratic fit including and excluding the static point. The conclusion is that
the quadratic fits give a smaller 
value for
\pagebreak 
the
physical B-parameters than the linear one
and are in better agreement with the static
value. Nevertheless, the predictions 
of all quadratic fits are still compatible
within 
 errors with the linear fit.

\end{document}